\documentclass[9pt,twocolumn,twoside]{pnas-new}

\templatetype{pnasresearcharticle} 

\begin{document}

\title{Universal scaling in far-from-equilibrium quantum systems: an equivalent differential approach}

\author[a]{Lucas Madeira}
\author[a]{Arnol D. Garc\'{i}a-Orozco}
\author[a]{Michelle A. Moreno-Armijos}
\author[a]{Amilson R. Fritsch}
\author[a,b,c,1]{Vanderlei S. Bagnato}

\affil[a]{Instituto de F\'isica de S\~ao Carlos, Universidade de S\~ao Paulo, CP 369, 13560-970 S\~ao Carlos, Brazil.}
\affil[b]{Department of Biomedical Engineering, Texas A\&M University, College Station, TX 77843, USA}
\affil[c]{Department of Physics \& Astronomy, Texas A\&M University, College Station, TX 77843, USA}

\leadauthor{Madeira}

\significancestatement{A challenging task has been understanding the route an out-of-equilibrium system takes to its thermalized state. Recent works indicate that some far-from-equilibrium systems display universal dynamics when close to a non-thermal fixed point (NTFP). In this work, we introduce a differential equation that has the universal scaling associated with NTFPs as a solution. The advantage of working with a differential equation, rather than only with its solution, is that we can extract several insightful properties not present in the solution alone. This equation can be applied to any system where NTFPs are present, but here we focus on studying Bose gases. The equation and its implications hold the potential to offer unexplored insights into the investigation of turbulence. }

\authordeclaration{The authors declare no competing interest.}
\correspondingauthor{\textsuperscript{1}To whom correspondence should be addressed. E-mail: vander@ifsc.usp.br}


\begin{abstract}
Recent progress in out-of-equilibrium closed quantum systems has significantly advanced the understanding of mechanisms behind their evolution towards thermalization. Notably, the concept of non-thermal fixed points (NTFPs) - responsible for the emergence of spatio-temporal universal scaling in far-from-equilibrium systems - has played a crucial role in both theoretical and experimental investigations. In this work, we introduce a differential equation that has the universal scaling associated with NTFPs as a solution. The advantage of working with a differential equation, rather than only with its solution, is that we can extract several insightful properties not necessarily present in the solution alone. How the differential equation is derived allows physical interpretation of the universal exponents in terms of the time dependence of the amplitude of the distributions and their momentum scaling. Employing two limiting cases of the equation, we determined the universal exponents related to the scaling using the distributions near just two momentum values. We established a solid agreement with previous investigations by validating this approach with three distinct physical systems. This consistency highlights the universal nature of scaling due to NTFPs and emphasizes the predictive capabilities of the proposed differential equation. Moreover, under specific conditions, the equation predicts a power-law related to the ratio of the two universal exponents, leading to implications concerning particle and energy transport. This suggests that the observed power-laws in far-from-equilibrium turbulent fluids could be related to the universal scaling due to NTFPs, potentially offering new insights into the study of turbulence.
\end{abstract}

\dates{This manuscript was compiled on \today}
\doi{\url{www.pnas.org/cgi/doi/10.1073/pnas.XXXXXXXXXX}}

\maketitle
\thispagestyle{firststyle}
\ifthenelse{\boolean{shortarticle}}{\ifthenelse{\boolean{singlecolumn}}{\abscontentformatted}{\abscontent}}{}

\firstpage[6]{4}


\dropcap{T}he field of out-of-equilibrium closed quantum systems encompasses many challenges, and recent times have witnessed remarkable progress in addressing many of these difficulties. A central pursuit in this field is untangling the mechanisms and routes responsible for taking non-equilibrium quantum systems towards thermalization. This topic is present in many areas of physics spanning all scales, from cosmological to particle physics systems~\cite{Polkovnikov2011}. Specifically, concepts such as pre-thermalization~\cite{Baier2001,Micha2003,Berges2004,Barnett2011,Nowak2014,Berges2014,Ueda2020} and non-thermal fixed points (NTFPs)~\cite{Scheppach2010,Berges2015} have contributed notably to advances in both theoretical understanding and experimental exploration of far-from-equilibrium systems.

Recently, a proposition has surfaced regarding classifying out-of-equilibrium quantum systems in universality classes~\cite{Schmidt2012,Nowak2013,Mikheev2023}. This concept parallels the notion of universality that emerges from fixed points observed in phase transition theories of systems in equilibrium. This dynamical analogue introduces the idea of NTFPs, which are metastable states within quantum many-body systems driven far from equilibrium. In proximity to these points, systems exhibit a notable independence of their initial conditions, with their dynamic evolution described by only a few parameters. These theoretical constructs are useful in interpreting various out-of-equilibrium phenomena more broadly~\cite{Nowak2012,Karl2013,Orioli2015,Schmied2019,Chantesana2019}.

Experiments involving cold atoms in the quantum regime have provided a unique platform for testing novel theories, such as the universal scaling in the neighborhood of NTFPs~\cite{Madeira2022}.  Particularly, compelling scenarios include quantum turbulence~\cite{Madeira2020} achieved in Bose-Einstein condensates (BECs)~\cite{Henn2009prl,Garcia2022} and quenches in cold atom gases~\cite{Karl2013_universal,Chiocchetta2015,Nicklas2015,Prufer2018,Eigen2018,Erne2018,Mikheev2019,Fujimoto2019,Jian2019,Glidden2021,Galka2022,Lannig2023}. These systems provide time-evolving quantities which can be measured and compared against theoretical predictions. 

This work presents compelling evidence that the phenomenon of universal scaling follows a well-defined differential equation, leading to the emergence of insightful properties. One can obtain accurate estimations for the exponents governing the universality class by examining two limiting cases. We demonstrate the validity of these principles through their application to three experimental investigations~\cite{Garcia2022,Prufer2018,Glidden2021}. Additionally, the equation predicts a power-law behavior near time-independent momentum distribution regions, and its characteristic exponent is related to the NTFP universal exponents.

\section*{NTFPs and universal scaling}

It has been suggested that out-of-equilibrium closed systems, falling within a specific universality category, manifest their self-similar behavior via their time-evolving momentum distribution $n(k,t)$~\cite{Schmidt2012,Nowak2013}, which is assumed to depend on the modulus $k=|\textbf{k}|$ of the momentum. When in the vicinity of a NTFP, this distribution obeys temporal and momentum scaling, adhering to the form:
\begin{equation}
\label{eq:scaling}
n(k,t)= \left(\frac{t}{t_0}\right)^\alpha F\left[\left(\frac{t}{t_0}\right)^\beta k \right],
\end{equation}
where $t_0$ is an arbitrary reference time (restricted to the temporal window where universal scaling is observed) and $F(k)$ is the corresponding universal function. This distinctive behavior emerges due to metastable regions within specific time intervals and momentum ranges and depends on two universal exponents. The $\alpha$ and $\beta$ exponents are related to amplitude adjustments and scale renormalizations, respectively. The traditional approach to investigating NTFPs involves perturbing the system from its equilibrium state and tracking the evolution of the momentum distribution $n(k,t)$.

\section*{Differential equation}

The standard method for identifying the presence of NTFPs involves verifying whether numerical or experimental data displays the universal scaling of Eq.~(\ref{eq:scaling}). This means examining if, for a specific time interval and momentum range, two exponents make all momentum distributions collapse into a single universal function $F(k)$. Taking a different approach, we can find additional properties nested within the non-equilibrium quantum system as it navigates its route toward thermalization.

We begin our exploration by focusing on an out-of-equilibrium system defined by the $n(k,t)$ distribution. If we consider the momentum as a function of time, $k=k(t)$, the total temporal derivative may be expressed as:
\begin{equation}
    \frac{dn}{dt}=\frac{\partial n}{\partial t}+\frac{\partial n}{\partial k}\frac{dk}{dt}.
\end{equation}
We must allow both amplitude and momentum scale variations to be time-parameterized to observe a self-similar scaling. Since the $\alpha$ exponent is responsible for rescaling the amplitude, we assume it is related to the total time derivative of the distribution. The $\beta$ exponent adjusts the momentum; thus, we associate it with the time dependence of $k(t)$. We consider a particular form for these time dependencies, which are power-laws in time,
\begin{eqnarray}
\label{eq:scale_n}
\frac{dn}{n}&=&\alpha\frac{dt}{t},\\
\label{eq:scale_k}
\frac{dk}{k}&=&-\beta\frac{dt}{t},
\end{eqnarray}
where $\alpha$ and $\beta$ are the exponents of such dependencies. Modest magnitudes of $\alpha$ and $\beta$ imply a slow evolution of the system, whereas the limiting case of $\alpha=\beta=0$ corresponds to a stationary state, indicating equilibrium.

With these assumptions, we conclude that $n(k,t)$ obeys the partial differential equation:
\begin{equation}
\label{eq:diff}
t\frac{\partial n(k,t)}{\partial t}=\alpha n(k,t)+\beta k \frac{\partial n(k,t)}{\partial k}.
\end{equation}
This equation contains the interplay between temporal changes in amplitude and momentum scale considering the power-law behaviors, Eqs.~(\ref{eq:scale_n}) and (\ref{eq:scale_k}), exhibited by both quantities. If $\alpha$ and $\beta$ are constants, a solution is Eq.~(\ref{eq:scaling}), which can be verified by direct substitution. Hence, this differential formulation gives an interpretation of the physical role of the universal exponents: choosing the particular time dependencies of Eqs.~(\ref{eq:scale_n}) and (\ref{eq:scale_k}) describe the phenomena responsible for this universal scalability. Different forms for Eqs.~(\ref{eq:scale_n}) and (\ref{eq:scale_k}) would lead to a distinct differential equation to which Eq.~(\ref{eq:scaling}) would not necessarily be a solution.

Equation~(\ref{eq:diff}) is not written in terms of the microscopic parameters of the system and, besides Eqs.~(\ref{eq:scale_n}) and (\ref{eq:scale_k}), we have also assumed the partial differential equation to be linear. In this sense, this equation may be viewed as phenomenological, and thus, it needs to be benchmarked with experimental data to show its correctness, which is one of the main goals of this work.

The advantage of working with a differential equation, rather than only with its solution, is that we can extract several insightful properties not necessarily present in Eq.~(\ref{eq:scaling}). It is worth noting that in the specific case of Boltzmann-type kinetic equations that display scaling solutions, integro-differential equations have been proposed~\cite{Chantesana2019}, where the time dependence of the momentum distribution is associated with a scattering integral. Equation~(\ref{eq:diff}) is more general in the sense that it describes any scaling solution regardless of its physical mechanism.

At this point, we can examine two limiting scenarios. The first case concerns momentum ranges where the amplitude of the momentum distribution undergoes negligible temporal variation, essentially characterizing regions of minimal change over time. Accordingly, the term $\partial n/\partial t$ can be neglected and Eq.~(\ref{eq:diff}) leads to:
\begin{equation}
\frac{\partial n}{\partial k}\Bigg|_t = -\frac{\alpha}{\beta} \frac{n}{k},
\end{equation}
yielding a solution applicable to this scenario of temporal stagnation,
\begin{equation}
\label{eq:limit_dn_dt_0}
n(k,t) \propto k^{-\alpha/\beta} \text{ (if $n$ is time-independent)},
\end{equation}
thus exhibiting a power-law behavior within this specific region.

\begin{figure*}[t]
\begin{center}
\includegraphics[width=\linewidth]{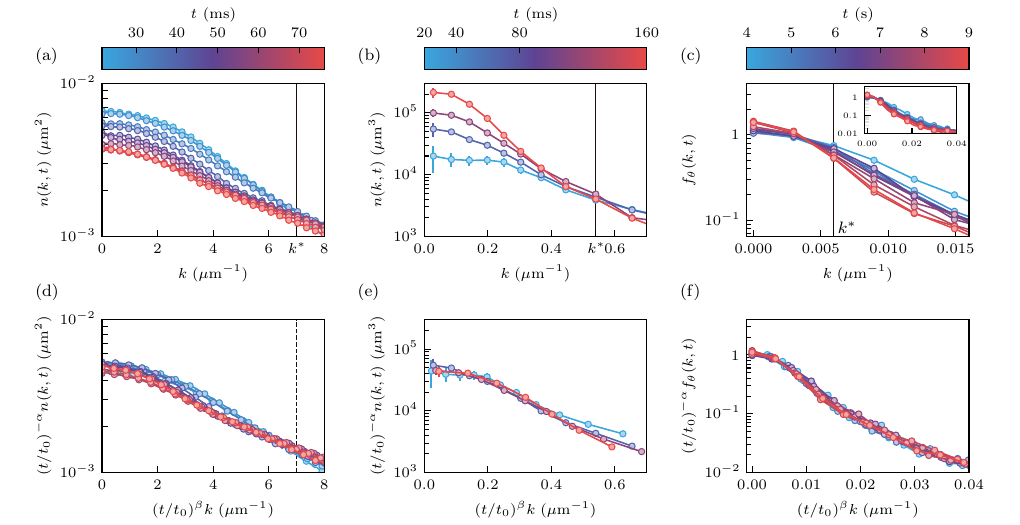}
\caption{Distributions and their collapse onto universal curves for three different physical systems. (a) Momentum distributions of a turbulent state of a harmonically trapped three-dimensional Bose gas for an excitation amplitude of $A=2\mu_0$ and (d) its scaling with $\alpha=-0.50(8)$ and $\beta=-0.2(4)$~\cite{Garcia2022}. The IR region in panel (d), where the universal dynamics are observed, extends from $k\approx 0$ until the dashed line.
(b) Momentum distribution for a quenched three-dimensional Bose gas, and (e) its corresponding IR scaling with $\alpha=1.15(8)$ and $\beta=0.34(5)$~\cite{Glidden2021,data_Glidden2021}. The error bars are taken from Ref.~\cite{data_Glidden2021}. (c) Structure factor of a spinor Bose gas driven far from equilibrium, where the inset shows a larger momentum range, and (f) the universal scaling with $\alpha=0.33(8)$ and $\beta=0.54(6)$~\cite{Prufer2018}.
Panels (a)-(c) indicate the momentum values $k^*=7.00, 0.54,$ and $0.006\mu$m$^{-1}$, respectively, where the distributions intersect.
For these three systems, universal scaling was observed in the IR region, which is limited by a momentum cutoff. For momentum values outside the IR region, the scaled distributions should not coincide, as seen in panels (d) and (e).
In all cases, the color bars denote the temporal evolution after driving the system far from equilibrium.
}
\label{fig:scaling}
\end{center}
\end{figure*}

For the particular case of wave turbulence~\cite{Nazarenko2011}, Svistunov~\cite{Svistunov1991} found turbulent cascades to be a result of $\partial n/\partial t=0$ (for a fixed momentum) and obtained a similar relation to Eq.~(\ref{eq:limit_dn_dt_0}). Another important point is that the region of minimal temporal variation means, considering the continuity equation~\cite{Orozco2020}, a momentum range where the flow of particles (or energy) also has a minimum variation with respect to $k$. In quantum turbulence, the constant flow with momentum characterizes the region of minimum dissipation and determines the so-called inertial region. This indicates that the turbulent cascade may be intrinsically associated with NTFP regions and mechanisms in the more general class of far-from-equilibrium quantum systems.

Indeed, this behavior of Eq.~(\ref{eq:limit_dn_dt_0}) is present implicitly in the universal function since by taking its derivative in time and equating it to zero, one obtains the same result. However, it is now evident that the power-law regime (with exponent $-\alpha/\beta$) occurs around the momentum regions where less temporal variation is expected. This provides a straightforward way to analyze the data. Instead of tracking $n(k,t)$, we may focus on $\partial n/\partial t$ for each $k$, and then it is possible to identify the regions of minimum variations and if they express power-law behavior. The exponents in such regions can indicate whether it is a phenomenon associated with NTFPs or another type.

Similarly, another interesting limit of Eq.~(\ref{eq:diff}) is $k\to 0$, deep in the infrared (IR) region. In this limit, Eq.~(\ref{eq:diff}) anticipates
\begin{equation}
\label{eq:limit_k_0}
n(k \to 0,t)\propto t^\alpha,
\end{equation}
consistent with taking the same limit directly in Eq.~(\ref{eq:scaling}). One of the advantages of Eq.~(\ref{eq:limit_k_0}) is that it makes tracking the temporal evolution of these non-thermal states more straightforward and practical. Instead of the whole $n(k,t)$ distribution, we can focus only on the low-momentum limit to obtain the $\alpha$ exponent. Following the amplitude of particles occupying the low-momentum modes enables us to observe the formation or destruction of the condensate and its relaxation as a function of time.

Equations~(\ref{eq:limit_dn_dt_0}) and (\ref{eq:limit_k_0}) show that when an out-of-equilibrium system approaches an NTFP, the exponents $\alpha$ and $\beta$ governing its evolution can be determined without the need to scale all curves to validate their collapse onto a universal curve.

\section*{NTFPs in experiments with cold atoms}

Many experimental investigations have employed cold atom gases as platforms to observe systems near NTFPs~\cite{Garcia2022,Nicklas2015,Prufer2018,Eigen2018,Erne2018,Glidden2021,
Galka2022,Lannig2023}. Here, we focus on three of them, which used very different physical systems and mechanisms to drive the system far from equilibrium. In Ref.~\cite{Garcia2022}, the authors investigated the emergence of universal scaling due to NTFPs in a harmonically trapped three-dimensional (3D) Bose gas driven to a turbulent state. A sinusoidal time-varying magnetic field gradient produced a far-from-equilibrium state. The amplitude $A$ of the excitation could be varied, and for three of its values, the authors showed that dynamical scaling emerges. Figure~\ref{fig:scaling}a shows the time evolution of the momentum distribution of a turbulent state. As time passes, the distribution shifts towards high momenta, indicating the depletion of the condensate. The scaling employing Eq.~(\ref{eq:scaling}) with $\alpha=-0.50(8)$ and $\beta=-0.2(4)$ is shown in Fig.~\ref{fig:scaling}d, which collapses all curves into a single function. These results are for an excitation amplitude of $A=2\mu_0$, where $\mu_0$ is the chemical potential at the center of the cloud. For the other amplitudes, the reader is referred to the Materials and Methods section.

\begin{figure*}[t]
\begin{center}
\includegraphics[width=\linewidth]{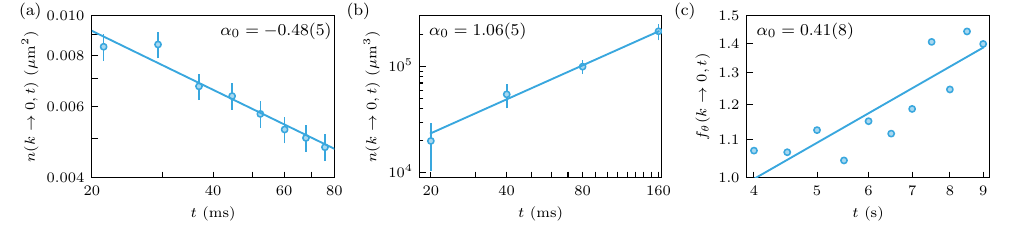}
\caption{Low-momentum behavior of the distributions as a function of time. (a) Turbulent BEC~\cite{Garcia2022}, (b) quenched Bose gas~\cite{Glidden2021}, and (c) spinor Bose gas~\cite{Prufer2018}. The solid curves correspond to a fit to the functional form provided by Eq.~(\ref{eq:limit_k_0}), which yield $\alpha$ values in agreement with the ones reported by the authors. The error bars are those reported in Refs.~\cite{Garcia2022,Glidden2021} and were not available for Ref.~\cite{Prufer2018}.
}
\label{fig:smallk}
\end{center}
\end{figure*}

Next, we consider Ref.~\cite{Glidden2021}, where the authors observed bidirectional dynamical scaling in a three-dimensional Bose gas~\cite{Gaunt2013}.
In this context, bidirectional means two flows with different directions: particles towards the IR region of the momentum distribution and energy towards the ultraviolet (UV) region. However, here, we will focus on the IR scaling.
Despite distinct exponents for universal scaling in each region, these values do not correspond to separate NTFPs; instead, they arise due to particle and energy transport following NTFP theory~\cite{Schmied2019}. The experimental strategy involved rapidly removing atoms and energy by turning off interatomic interactions and lowering the trap depth, inducing a far-from-equilibrium state. Turning on the interactions initiated thermalization. Figure~\ref{fig:scaling}b, produced with the data available in Ref.~\cite{data_Glidden2021}, illustrates unscaled $n(k,t)$ profiles, while the IR region displays universal behavior with $\alpha=1.15(8)$ and $\beta=0.34(5)$, as seen in Fig.~\ref{fig:scaling}e.

The third system we consider was investigated in Ref.~\cite{Prufer2018}, where authors observed universal dynamics using a quasi-one-dimensional spinor Bose gas~\cite{Sadler2006}. This was achieved through an analysis of spin correlations in a spin-1 system. With the initial state corresponding to all atoms occupying a single magnetic state ($m_F=0$), the departure from equilibrium was induced by an abrupt alteration in the energy splitting of the $F=1$ sublevels, thus producing excitations in the $F_x-F_y$ spin plane.
The authors computed the mean spin length on the $F_x-F_y$ plane, a local angle $\theta$ was defined, and the fluctuations were measured using a two-point correlation function. The function entering Eq.~(\ref{eq:scaling}) is the averaged Fourier transform of the correlation function, called the structure factor $f_\theta(k,t)$. Although this is not the momentum distribution of the system, as was the case for the two previous examples, this function also depends on momentum and time.
In Fig.~\ref{fig:scaling}c, we reproduce the temporal evolution of the structure factor $f_\theta(k,t)$ using the data provided by the authors~\cite{Prufer2018}. There is a discernible shift towards lower momenta as time passes. Applying the scaling of Eq.~(\ref{eq:scaling}) with $\alpha=0.33(8)$ and $\beta=0.54(6)$ to the data results in the convergence of all points onto a universal curve, as depicted in Fig.~\ref{fig:scaling}f.

Although other experiments involving NTFPs have been performed, these three were chosen in this work because they illustrate very different dynamics in cold bosonic gases. The values of the exponents are positive in Refs.~\cite{Glidden2021,Prufer2018}, since there is a particle transport towards the IR region, while the excitation protocol in Ref.~\cite{Garcia2022} is responsible for a particle flow towards the UV region, which yields negative exponents. Moreover, the distinct physical mechanisms of driving the gases to far-from-equilibrium states have different time scales: the excitation in Ref.~\cite{Garcia2022} and the quench in Ref.~\cite{Glidden2021} display universal behavior over tens of milliseconds, while the spinor dynamics~\cite{Prufer2018} takes seconds.

\section*{Results}

References~\cite{Garcia2022,Glidden2021,Prufer2018} employed statistical analyses to obtain the universal exponents $\alpha$ and $\beta$ that collapse the distributions for a momentum range and time interval onto a single function. These procedures employ all the experimental points in the considered interval. However, if the scaling interval contains the $k\approx 0$ region, Eq.~(\ref{eq:limit_k_0}) is a much simpler way of determining $\alpha$. In Fig.~\ref{fig:smallk}, we show the time evolution of each distribution for the smallest momentum measured. A fit to the functional form of Eq.~(\ref{eq:limit_k_0}) provides an estimate of the value of $\alpha$, in agreement with the ones found in Refs.~\cite{Garcia2022,Glidden2021,Prufer2018}. We summarize the results obtained for $\alpha$ in the first two columns of Tab.~\ref{tab:alpha_beta}.

\begin{table*}[t]
\caption{Universal exponents. The columns ``standard procedure'' correspond to the $\alpha$ and $\beta$ values reported by the authors of Refs.~\cite{Garcia2022,Glidden2021,Prufer2018}. In the case of Ref.~\cite{Garcia2022}, three different excitation amplitudes were employed. ``Taking the $k\to 0$ limit'' stands for the low-momentum approximation, see Fig.~\ref{fig:smallk}. Equation~(\ref{eq:beta_kstar}) was used to compute the values $\beta_0$ and $\beta_1$, the difference being which value of $\alpha$ was employed, hence the labels of the last two columns.
}
\vspace{0.2cm}
\begin{tabular}{cc||c|c||c|c|c}
\multicolumn{2}{c||}{} & Standard procedure & Taking the $k \to 0$ limit & Standard procedure & Near $k^*$ with $\alpha_0$ & Near $k^*$ with $\alpha$ \\
\multicolumn{2}{c||}{System} & $\alpha$	& $\alpha_0$ & $\beta$ & $\beta_0$     & $\beta_1$ \\ \hline\hline
&A ($\mu_0$)&&&&& \\
&1.8  		& -0.46(2) 	& -0.44(5)   & -0.2(9)	& -0.26(7)   	& -0.30(7) \\ 
Ref.~\cite{Garcia2022} &2.0  		& -0.51(1) 	& -0.48(5)   & -0.2(7)	& -0.25(5)   	& -0.26(5) \\ 
&2.2  		& -0.50(2)  & -0.49(5)   & -0.2(9)	& -0.37(8)   	& -0.38(9) \\ \hline
\multicolumn{2}{c||}{Ref.~\cite{Glidden2021}}			& 1.15(8)	& 1.06(5)	 & 0.34(5)	& 0.41(7)		& 0.45(7)	\\ \hline
\multicolumn{2}{c||}{Ref.~\cite{Prufer2018}}			& 0.33(8)	& 0.41(8)	 & 0.54(6)	& 0.57(9)		& 0.46(7)
\end{tabular}
\label{tab:alpha_beta}
\end{table*}

Having determined $\alpha$ from the $k\to 0$ behavior of the distributions, we now focus on obtaining the value of $\beta$. Figure~\ref{fig:scaling} shows that the unscaled curves for the three physical systems all possess a momentum value where the distribution is approximately time-independent, which we denote by $k^*$ (for the harmonically-trapped BEC, this is more easily identified in Fig.~\ref{fig:kstar} of the extended data section).
In the vicinity of this point, Eq.~(\ref{eq:diff}) can be used to determine the value of $\beta$,
\begin{equation}
\label{eq:beta_kstar}
\beta = -\frac{\alpha}{k^*} \frac{n(k^*,t)}{(\partial n/\partial k)|_{k=k^*}}.
\end{equation}
We employed Eq.~(\ref{eq:beta_kstar}) to compute the values of $\beta$ for the three physical systems of interest. Since Eq.~(\ref{eq:beta_kstar}) requires a value for $\alpha$, we considered two cases: the value found with the $k\to 0$ behavior of the distributions and the one reported by the authors of Refs.~\cite{Garcia2022,Glidden2021,Prufer2018}. The results are summarized in Tab.~\ref{tab:alpha_beta}, showing that our differential approach reproduces the same values of $\beta$ obtained using the standard procedure.

In the case of Ref.~\cite{Garcia2022}, the relevant momentum values for the universal scaling are within a restricted range of 0 to $\approx$ 10 $\mu$m$^{-1}$, while in this range, the amplitude of $n(k,t)$ varies less than a logarithmic scale. This limited range of quantities is inherent to the system and makes the experimental errors larger since their evaluation comes from averages and analysis of distribution functions. Since the scalability in $k$ is related to $\beta$, and this scale is the most limited in range, it is expected to be a more significant error in this exponent than in $\alpha$, as it can be seen in the uncertainties reported using the standard procedure in Table~\ref{tab:alpha_beta}. One of the advantages of the method we proposed in this manuscript is that it yields uncertainties one order of magnitude smaller because the value of the momentum distribution (and its momentum derivative) are evaluated only at one momentum point ($k^*$) to estimate $\beta$.

\begin{figure}[!htb]
\begin{center}
\includegraphics[width=\linewidth]{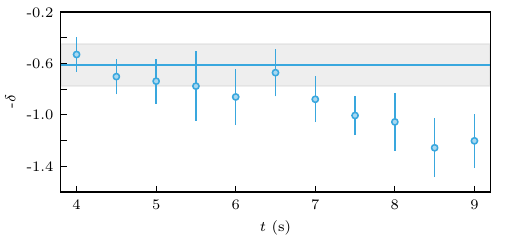}
\caption{
Exponent of the power-law as a function of the time. The functional form $f_\theta(k,t)\propto k^{-\delta}$ is fitted near $k^*\approx 0.006 \; \mu$m$^{-1}$ for the case of a spinor Bose gas~\cite{Prufer2018}. The error bars correspond to the uncertainty due to the fitting procedure, see the Materials and Methods section. The theoretical prediction of Eq.~(\ref{eq:limit_dn_dt_0}) is $\delta=\alpha/\beta=0.61(16)$, which is denoted by the horizontal blue line and its uncertainty by the shaded area.
}
\label{fig:exponent}
\end{center}
\end{figure}

Finally, Eq.~(\ref{eq:limit_dn_dt_0}) predicts a power-law for $n(k,t)$ in a momentum range where the distribution is approximately time-independent. Inspection of the unscaled distributions in Fig.~\ref{fig:scaling} reveals that, although we can identify a point $k^*$ where $(\partial n/\partial t)|_{k=k^*}\approx 0$, this condition is not rigorously observed in a finite $k$-range for none of the three physical systems considered here. However, the beginning of the time evolution ($t\lesssim 6.5\;$s) of the spinor Bose gas fulfils this condition in good approximation for the range of $0.003\; \mu$m$^{-1}\leqslant k \leqslant 0.009\; \mu$m$^{-1}$. Figure~\ref{fig:exponent} shows the exponent obtained when we fitted the data in this region to a power-law functional form, $f_\theta(k,t)\propto k^{-\delta}$. We also included in the figure a horizontal line and a shaded region corresponding to the theoretical prediction of Eq.~(\ref{eq:limit_dn_dt_0}), $\delta=\alpha/\beta=0.61(16)$, calculated with the values and uncertainties reported in Ref.~\cite{Prufer2018}. The agreement is good for the early evolution ($t\lesssim 6.5\;$s), but for later times, the values of the structure factor change considerably, such that the premise of the theoretical prediction is inapplicable beyond this temporal threshold.

\section*{Discussion and Conclusions}

In this work, we proposed a differential equation, Eq.~(\ref{eq:diff}), which has as a solution the well-known scaling due to the presence of NTFPs, Eq.~(\ref{eq:scaling}). This allowed us to investigate features not evident or present in the solution alone. We then applied these properties to three distinct physical systems~\cite{Garcia2022,Glidden2021,Prufer2018} and found that our predictions agree with previous investigations. The fact that the systems and the protocols for driving them out of equilibrium are so different stresses the universal character of the scaling due to NTFPs and the results from our differential equation.

First, from the $k\to 0$ behavior of the distributions, we obtained an estimate for the universal exponent $\alpha$. Although this considers only one data point for each time instant, which is undesirable when working with experimental data, the agreement with the values reported in Refs.~\cite{Garcia2022,Glidden2021,Prufer2018} is remarkable.

Then, we identified a point, denoted by $k^*$, at which the temporal variation of the distributions is approximately null. This feature is present in the physical systems under investigation in this work and several other systems where universal scaling was observed~\cite{Nicklas2015,Eigen2018,Erne2018,Galka2022,Lannig2023}. Using the differential equation and the variation of the distributions with respect to momentum, we estimated the values of $\beta$. Hence, with information about the distributions at the vicinity of just two momentum values, $k\to 0$ and $k^*$, we were able to compute the universal exponents $\alpha$ and $\beta$ in agreement with the values obtained through much more sophisticated procedures~\cite{Garcia2022,Glidden2021,Prufer2018}.

In all physical systems, the universal dynamics of the distributions due to nearby NTFPs are expected to appear in finite time and momentum intervals due to the typical scales of the system. Finding these ranges using the universal scaling, Eq.~(\ref{eq:scaling}), requires considering all possible combinations of the available data to obtain the universal region, which can be a daunting task. Instead, if we consider the differential formulation, we can perform our analysis with the variations of the distributions with respect to momentum and time. Then, the time and momentum ranges where scalability appears are readily available: they correspond to the intervals where Eq.~(\ref{eq:diff}) is satisfied. This makes analyzing the data of experiments or simulations straightforward.

Equation~(\ref{eq:limit_dn_dt_0}) contains an interesting prediction: if the distribution is time-independent in a specific momentum range, then a power-law $\propto k^{-\alpha/\beta}$ is expected. Although the assumption of a constant distribution in a range is not strictly observed in the physical systems we considered, the early evolution of the spinor Bose gas agrees with the theoretical prediction reasonably.

Finally, another intriguing feature of Eq.~(\ref{eq:limit_dn_dt_0}) is that its power-law is related to the ratio of the two universal exponents, which has physical meaning under specific conditions. For that, let us define two quantities inside the region where scaling is observed, the mean particle number $\overline{N}$ and kinetic energy $\overline{E}$, which help us describe particle and energy transport. Using Eq.~(\ref{eq:scaling}), we can write the temporal dependence of these quantities in terms of the dimension $d$ of the system and universal exponents,
\begin{eqnarray}
\label{eq:nbar}
\overline{N}(t)
&=&\int 
d^dk \ n(k,t)
\propto \left(\frac{t}{t_0}\right)^{\alpha-d\beta},\\
\label{eq:ebar}
\overline{E}(t)& \propto &\int 
d^dk \ k^2 n(k,t) \propto \left(\frac{t}{t_0}\right)^{\alpha-(d+2)\beta},
\end{eqnarray}
where we assumed a quadratic dispersion relation to compute $\overline{E}$. The integrals are over the scaled range ${\left(t/t_0\right)^{-\beta}k_\ell \leqslant |k|\leqslant\left(t/t_0\right)^{-\beta}k_h}$, and $k_\ell\leqslant k \leqslant k_h$ is the momentum range where the universal scaling is observed.

According to Eq.~(\ref{eq:nbar}), particle-conserving transport requires $\alpha=d\beta$, while Eq.~(\ref{eq:ebar}) dictates that energy-conserving transport is given by $\alpha=(d+2)\beta$~\cite{Chantesana2019,Schmied2019}. Consequently, for these two types of conserving transports, $\alpha$ and $\beta$ assume the same sign, which is determined by the direction of the transport: positive towards the IR region and negative towards the UV region. Distinct momentum regions may exhibit different values for $\alpha$ and $\beta$, accompanied by different signs~\cite{Glidden2021}.
The theory of NTFPs comes from an analogy with the renormalization group (RG) theory in equilibrium critical phenomena~\cite{Mikheev2023}, where the time in the universal scaling due to NTFPs plays the role of space in equilibrium critical phenomena. Hence, it is possible to draw parallels with the equations that relate the universal exponents, dimensionality, and the dispersion relation with the scaling laws of the critical exponents near phase transitions in equilibrium. An example of such similarity is the Josephson hyperscaling relation~\cite{Kempkes2016,Arouca2020}, which merits future investigation to see to what extent the analogy holds. Taking advantage of the parallel with RG theory in critical phenomena, it would also be interesting to see how the approach introduced in this work relates to RG flow equations~\cite{Mikheev2023}.

Hence, Eq.~(\ref{eq:limit_dn_dt_0}) predicts a power-law $\propto k^{-d}$ for particle-conserving transport, while energy-conserving transport corresponds to $\propto k^{-(d+2)}$. These two situations are very relevant to the field of quantum turbulence~\cite{Henn2009prl,Madeira2020}, which may help to elucidate open questions. For example, during the pre-steady state of wave turbulence in a two-dimensional homogeneous Bose gas~\cite{Galka2022}, NTFP-like dynamic scaling was observed with $\alpha = \gamma \beta$, where $\gamma=2.90(5)$ is not $d=2$. Besides NTFP results, the theory of weak-wave turbulence~\cite{Nazarenko2011} also predicts $\gamma=d=2$ for this system. Hence, the origin of this difference remains to be investigated.

A standard approach to identify a turbulent state in a quantum fluid is the emergence of a power-law in the momentum distribution~\cite{Thompson2014,Navon2016}, indicating a particle cascade, or in the energy spectrum, a signal of an energy cascade. The differential analysis allows us to predict regions of power-law behavior, creating the possibility of determining the ratio of the exponents independently. Moreover, recently, it has been suggested that there are three main types of quantum turbulence (QT)~\cite{Barenghi2023}: Kolmogorov QT, Vinen QT, and strong QT. The physical mechanisms and several other features differ between them. However, they all display a power-law in some characteristic momentum range. Our result suggests that the power-laws observed in turbulent fluids, which are far from equilibrium, could be intimately related to the universal scaling due to NTFPs. These results bring a new perspective to investigate the complex phenomenon of turbulence and merit further investigation.

\matmethods{
\subsection*{Determining the exponents and their uncertainty}

We used a non-linear least squares procedure to fit the data to the function to determine the $\alpha$ exponents from the $k\to 0$ behavior of the distributions. The optimal values of the parameters are such that the sum of the squared residuals is minimized. We obtained the variances of the parameter estimates from the diagonal of the covariance matrix. The uncertainties reported in the main text and the Extended Data correspond to one standard deviation: the square root of the corresponding diagonal element of the covariance matrix.

For the $\beta$ exponents, obtained through Eq.~(\ref{eq:beta_kstar}), the first step is to determine $k^*$. This is achieved by finding the momentum value for which $n(k^*,t)$ has the smallest dispersion considering all values of $t$. The partial derivative of the distribution with respect to the momentum has to be taken numerically, which can lead to large fluctuations when applied to experimental data. Using a  central difference derivative, we found that the data of Ref.~\cite{Prufer2018} had to be smoothed to reduce the noise, and a three-point moving average was sufficient to prevent the fluctuations. Note that Eq.~(\ref{eq:beta_kstar}) yields a value of $\beta$ for each instant $t$; therefore, we report the average of the values and their uncertainty as one standard deviation.

For the $\delta$ exponent of the power-law, $f_\theta(k,t)\propto k^{-\delta}$,
we performed a non-linear least square fit for each instant $t$ in a momentum range including $k^*\approx 0.006 \; \mu$m$^{-1}$. The procedure and the error estimates are the same as the ones described above for determining $\alpha$.

\subsection*{Extended data}

To better illustrate the $k^*$ point, defined through $(\partial n/\partial t)|_{k=k^*}=0$, in Fig.~\ref{fig:kstar} we provide the same momentum distributions as in Fig.~\ref{fig:scaling}(a), but for a larger momentum range. This feature is present in harmonically trapped three-dimensional Bose gases driven to a turbulent state~\cite{Garcia2022}, quenched three-dimensional Bose gases~\cite{Glidden2021}, quasi-one-dimensional spinor Bose gases~\cite{Prufer2018}, and many other experiments where the systems are driven to far-from-equilibrium states.

\begin{figure*}[b]
\begin{center}
\includegraphics[width=\linewidth]{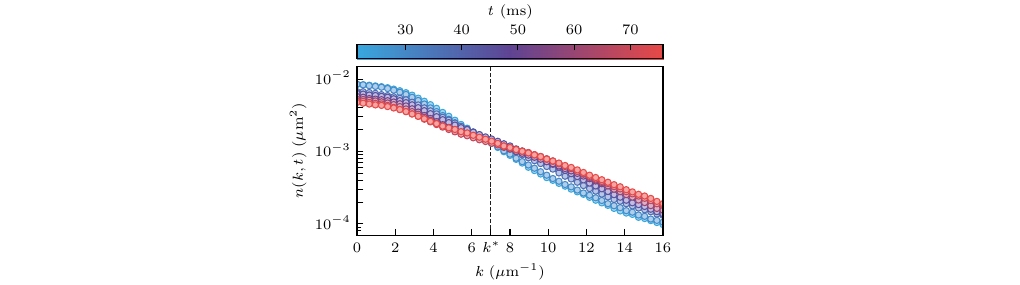}
\caption{
Momentum distributions of Fig.~\ref{fig:scaling}(a), but for a larger momentum range. The point $k^*$ is the momentum value where all distributions cross, indicating a negligible temporal variation of $n(k^*,t)$.
}
\label{fig:kstar}
\end{center}
\end{figure*}

In the main text, we provided the results for one of the excitation amplitudes of Ref.~\cite{Garcia2022}. Here, we also provide similar figures for the other two amplitudes. Figure~\ref{fig:scaling_sup} contains the distributions and their scaling, while Fig.~\ref{fig:smallk_sup} is related to the $k\to 0$ behavior of the distributions.

\begin{figure*}[!htb]
\begin{center}
\includegraphics[width=\linewidth]{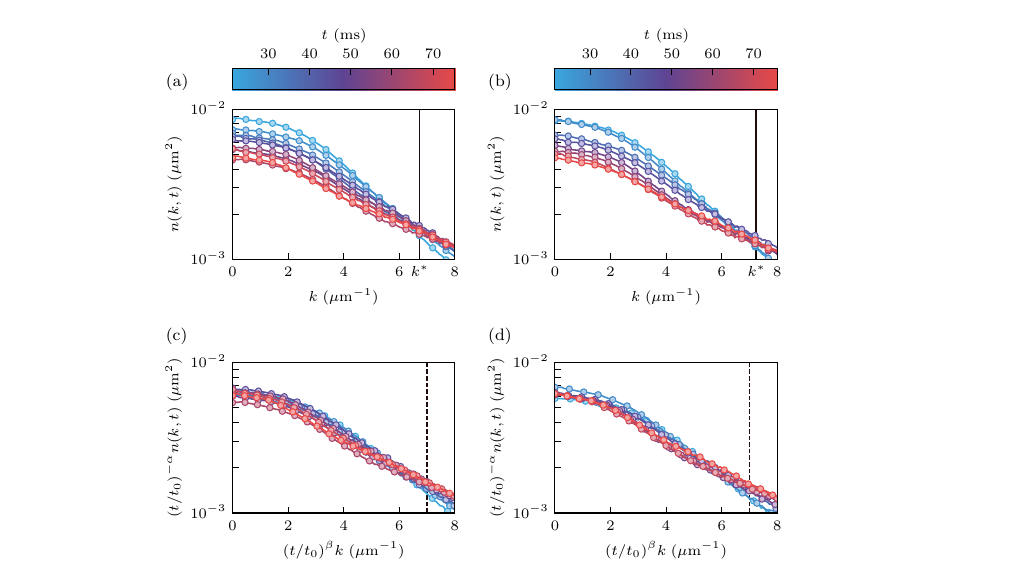}
\caption{
Momentum distributions of harmonically trapped three-dimensional Bose gases excited towards turbulence and their collapse onto universal curves.
The momentum distributions for two excitation amplitudes are shown in (a) and (b), $A=1.8\mu_0$ and $A=2.2\mu_0$, respectively,
which intersect at the momentum values $k^*=6.72$ and $7.24\mu$m$^{-1}$, respectively.
The universal scaling with $\alpha=-0.50(8)$ and $\beta=-0.2(4)$~\cite{Garcia2022} is displayed for the respective amplitudes in (c) and (d), where the dashed lines denote the upper limit of the IR scaling region.
}
\label{fig:scaling_sup}
\end{center}
\end{figure*}

\begin{figure*}[!htb]
\begin{center}
\includegraphics[width=\linewidth]{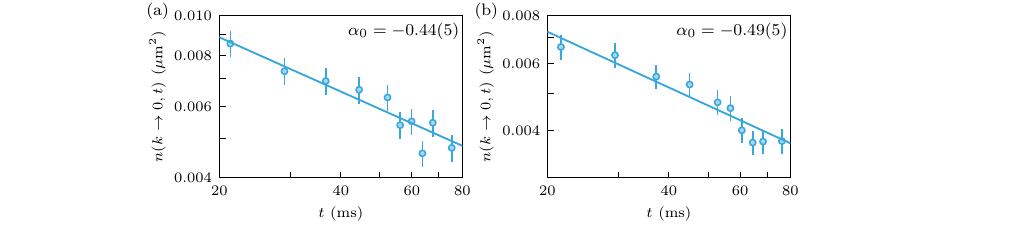}
\caption{Low-momentum behavior of $n(k,t)$ for turbulent states of harmonically trapped three-dimensional Bose gases as a function of time. (a) For excitation amplitudes of $A=1.8\mu_0$ and (b) $A=2.2\mu_0$. The solid curves correspond to a fit to the functional form provided by Eq.~(8).}
\label{fig:smallk_sup}
\end{center}
\end{figure*}

\subsection*{Data Availability}
Data related to the harmonically trapped three-dimensional Bose gas driven to a turbulent state is available at~\cite{Garcia2022} and as Supporting Information. Data concerning the bidirectional dynamical scaling in a three-dimensional Bose gas is available in Refs.~\cite{Glidden2021,data_Glidden2021}, while for the quasi-one-dimensional spinor Bose gas, the reader is referred to Ref.~\cite{Prufer2018}.
}

\showmatmethods{} 

\acknow{We thank T.~Gasenzer, Z.~Hadzibabic, and S.~Nazarenko for fruitful discussions. This work was supported by the S\~ao Paulo Research Foundation (FAPESP) under the grants 2013/07276-1, 
2014/50857-8, 
2022/00697-0, 
and 2023/04451-9, 
and by the
National Council for Scientific and Technological Development (CNPq)
under the grants 465360/2014-9 and 
381381/2023-4. 
M.A.M-A. acknowledges the support from Coordenação de Aperfeiçoamento de Pessoal de Nível Superior - Brasil (CAPES) - Finance Code 88887.643259/2021-00. 
}

\showacknow{} 

\bibsplit[2]

\bibliography{references}

\end{document}